# Theoretical Analysis of Quality of Conventional Beamforming for Phased Microphone Arrays

Dheepak Khatri[a] and Kenneth Granlund[a,1]

[a] *North Carolina State University, Raleigh, 27695, NC, USA*

**Abstract**

A theoretical study is performed to analyze the directional response of different types of microphone array designs. 1-D (linear) and 2-D (planar) microphone array types are considered, and the *delay and sum beamforming* and *conventional beamforming* techniques are employed to localize the sound source. A non-dimensional parameter, *G*, is characterized to simplify and standardize the rejection performance of both 1-D and 2-D microphone arrays as a function of array geometry and sound source parameters. This parameter *G* is then used to determine an improved design of a 2-D microphone array for far-field sound localization. One such design, termed the *Equi-area array* is introduced and analyzed in detail. The design is shown to have an advantageous rejection performance compared to other conventionally used 2-D planar microphone arrays.

---

[1] Email address: kgranlu@ncsu.edu

1. **Introduction**

It is often desired to receive sound from a particular direction or location, rejecting sound from unwanted directions or locations (noise). Using a single omnidirectional microphone would not be beneficial as it would capture sound and noise equally. A solution to this problem is sought in the form of directional microphones, which capture sound within a band of incoming angles, termed 'band of acceptance', effectively a cone. However, such types of microphones are not dynamic, i.e., a physical movement of the microphone is required to shift the direction of sound reception. Also, once manufactured, the 'band of acceptance' cannot be modified.

An alternative solution to this problem suggests the use of multiple omnidirectional microphones placed in an array format [1]. The fundamental theory behind this solution is that sound is a wave, and waves originating from different locations travel different distances, and arrive at the receiver (microphone) with a phase difference. By efficient design, it is possible to reject sound (noise) from certain locations by means of destructive interference (180° phase difference) and accept only sound from desired locations (constructive interference).

Such a technique of using signal processing to achieve directivity of sound is called Beamforming. A review of the topic is covered by Chiarotti [2]. This technique has been used for many years in various fields such as design of aircraft airframes and engines [3], design of wind turbine blades, wind tunnel tests for rotorcraft analysis, acoustic modeling of auditoriums, etc. and has expanded its horizons into new technologies such as speech acquisition tools [4], etc. The authors in specific are excited over the developments of this technique in the design of phased microphone arrays for airframe and undercarriage noise measurements. The motivation for this project is the development of an instrument that can measure localized sound production from canonical sources such as a vibrating disc or dilating sphere, while at the same time the local flow field is measured with Particle Image Velocimetry, to further understand the combined aero-acoustic effects. Here we aim to build a simplistic theoretical approach towards 1-D and 2-D beamforming, and qualitatively demonstrate on-axis performance of ~~different~~ array designs in both near- and far- field with a proposed better design than the popular spiral- or concentric ones. For a more thorough coverage of acoustic arrays; Merino-Martinez, et al. [5] provide a comprehensive review of different beamforming techniques; Sarrajd [6], Luesuttihiviboon et al. [7] and Rabinkin et al. [8] covers optimization of microphone array arrangements; Van Trees [9] covers the topic of array design as well as signal processing. Optimization of post-processing techniques for acoustic data, such as DAMAS [10] can in greatly increase the performance of acoustic arrays, especially in complex setups with high aerodynamic shear flows between the transmitter and receivers, but it is not covered in this work because of the high computational effort. For the proposed application of a vibrating disk, CLEAN-SC [11] is preferred according to Porteous et al. [12] because of the dipole source. All processing of data in this paper was done in Matlab [13] on a desktop computer.

## 2. Beamforming Algorithm

### 2.1 1-D Linear Array beamforming

In this setup, $N$ number of microphones are placed in a linear fashion, with equidistant spacing $d$. The incoming sound wave is assumed to be a planar wave, having an azimuth angle, $\theta$, with the line of the microphones. The side view of the arrangement can be seen in Fig. 1.

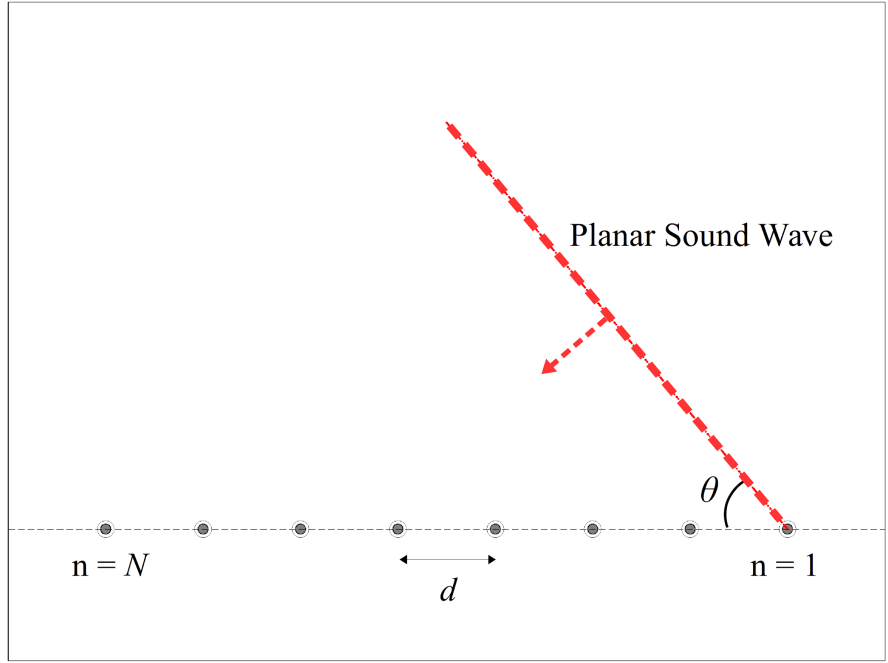

**Fig. 1.** A linear array with incoming sound.

The numbering of microphones starts from the right microphone. The sound wave direction is characterized by just one parameter, the azimuth angle, θ. Assuming a distortionless medium and that the sound wave arrives at the first microphone with zero phase difference, the phase difference $\phi_i$ observed by the $i$th microphone is given in Eq (1).

$$\phi_i = (i-1)\frac{2\pi f}{c} d \sin\theta \ (rad) \ \forall \ i = 1, 2 \ldots N \tag{1}$$

Where $f$ is the frequency of the sound wave in Hz, and $c$ is the speed of the sound wave in m/s.

### 2.1.1  Computing array response

Let $\theta_o$ be the desired direction of reception of sound. Therefore, the phase difference experienced by the $i^{th}$ microphone, for the sound incoming from $\theta_o$ direction is expressed in Eq (2).

$$\phi_{o_i} = (i-1)\frac{2\pi f}{c} d \sin\theta_o \tag{2}$$

Let $v_i(x)$ be the waveform of the soundwave incoming at the $i$th microphone from the desired direction. Beamforming is achieved by summing and averaging the received input at each microphone, but shifted by the required phase, $\phi_{o_i}$.

$$v_{signal} = \frac{[v_1(x-\phi_{o_1}) + v_2(x-\phi_{o_2}) + \cdots \ldots + v_N(x-\phi_{o_N})]}{N} \tag{3}$$

If the sound wave was incoming only from the desired direction, each of the individual elements of $v_{signal}$ would add up perfectly, as we are subtracting the exact phase difference that was caused due to excess distance travelled. However, this is not always the case. In practice, there is noise coming in from every other direction θ. Therefore, from the same beamforming technique, the noise that will captured, coming in from direction $\theta \neq \theta_o$ is:

$$v_{noise} = \frac{[v_1(x+\phi_1-\phi_{o_1})+v_2(x+\phi_2-\phi_{o_2})+\cdots+v_N(x+\phi_N-\phi_{o_N})]}{N} \quad (4)$$

This type of Beamforming technique is termed as *Delay and Sum Beamforming* [14]. The above methodology is its application for a 1D array, and is of ~ O(*N*), as the phase difference of each microphone is computed with respect to a single microphone (the first microphone in this case), and the beamforming for $v_{signal}$ and $v_{noise}$ contains the sum of *N* elements.

We are mostly interested in the amplitude of the sound recorded, and not the entire waveform. Therefore, all array response parameters are computed with respect to the amplitude of the sound and normalized with the amplitude of the sound from the desired direction.

We are interested in defining parameters that help us in understanding the efficiency of the beamforming algorithm and the phased microphone array design. The beamforming is considered ideal if it is capable of eliminating all noise, and ineffective if the signal and noise are equal and indistinguishable. This takes the following mathematical form,

$$\eta(\theta) = \left|\frac{v_{signal} - v_{noise}(\theta)}{v_{signal}}\right| = 1 - \frac{|v_{noise}(\theta)|}{|v_{signal}|} \quad (5)$$

The overall rejection of noise from all azimuth angles ($\theta$), can be obtained by evaluating an average integral of the noise and signal beamforming components. This is termed as the Rejection Factor ($RF$).

$$RF = 1 - \left[\frac{\int_{\theta=-\frac{\pi}{2}}^{\theta=\frac{\pi}{2}} |v_{noise}(\theta)|(d\theta)}{\int_{\theta=-\frac{\pi}{2}}^{\theta=\frac{\pi}{2}} |v_{signal}|(d\theta)}\right] = 1 - \left[\frac{1}{\pi}\int_{\theta=-\frac{\pi}{2}}^{\theta=\frac{\pi}{2}} \frac{|v_{noise}(\theta)|}{|v_{signal}|}(d\theta)\right] \qquad (6)$$

The value of the rejection factor always lies between 0 and 1, with 1 being complete rejection of noise, and 0 being no rejection of noise. The rejection factor can also be expressed in terms of a percentage, termed as Rejection Percentage ($RP$) in the following sections.

The Signal-to-Noise ratio (in dB) is defined as

$$SNR\ (dB) = 20\ log_{10}\left[\frac{\int_{\theta=-\frac{\pi}{2}}^{\theta=\frac{\pi}{2}} |v_{signal}|(d\theta)}{\int_{\theta=-\frac{\pi}{2}}^{\theta=\frac{\pi}{2}} |v_{noise}(\theta)|(d\theta)}\right] \qquad (7)$$

From Eq (6) and (7), the SNR (dB) can be expressed in terms of $RF$ as,

$$SNR\ (dB) = -20\ log_{10}([1 - RF]) \qquad (8)$$

The performance of an array could be improved by employing what is termed *Conventional Beamforming* [15]. In this beamforming technique, the phase difference of each microphone is computed with respect to every other microphone, thereby forming a ($N \times N$) matrix of phase differences, where each ($i,j$) term represents the phase difference of microphone $i$, with respect to microphone $j$.

$$\phi_{ij} = (i-j)\frac{2\pi f}{c} d \sin\theta \tag{9}$$

Therefore, the beamforming for $v_{signal}$ and $v_{noise}$ using this technique is of ~ O($N^2$) as it contains the sum of $N^2$ elements.

*2.2    2-D planar Array beamforming*

In this setup, $N$ number of microphones are placed on a planar disk in the required design.

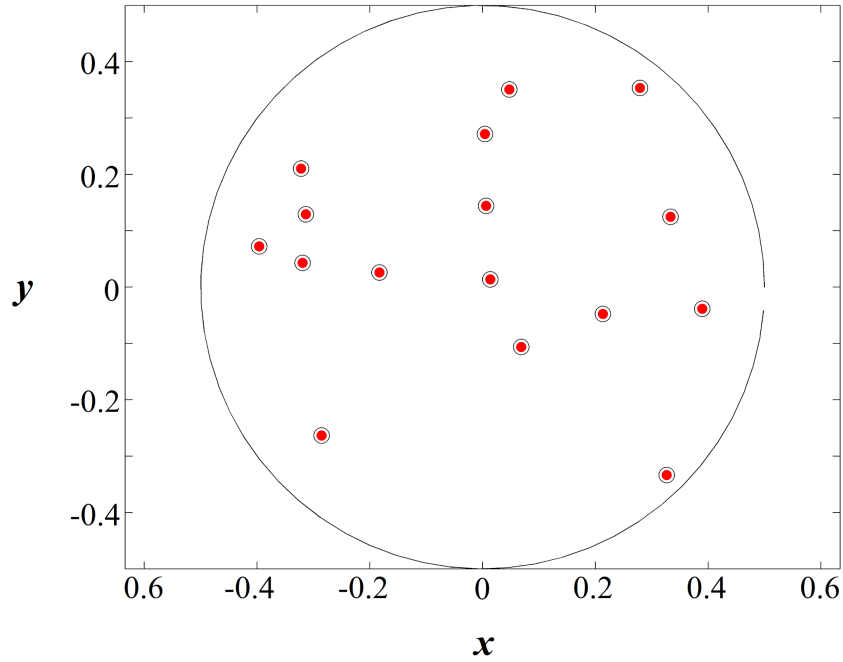

**Fig. 2.** Microphones positioned randomly on a planar disk.

Once again conventional beamforming ($\sim O(N^2)$) technique is employed. Consider any two microphones belonging to the array. Let the positions of the microphones using cartesian coordinates be $(x_1, y_1, 0)$ and $(x_2, y_2, 0)$. Let the position of an arbitrary sound source be $(x_s, y_s, z_s)$. Since the frame of reference lies in the plane of the microphones, $z_1 = z_2 = 0$.

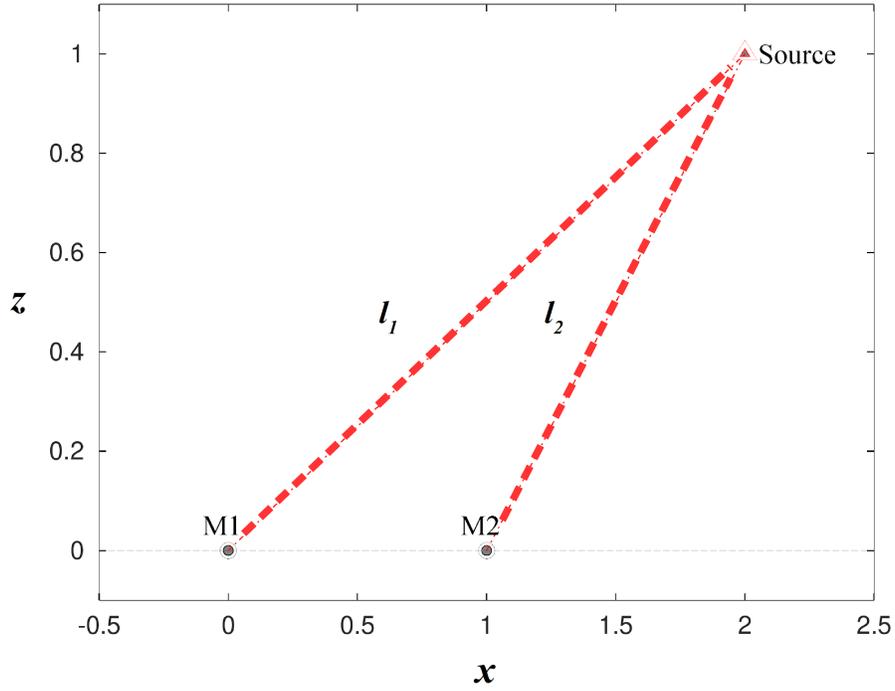

**Fig. 3.** Two microphones M1 and M2 and point source system

Let $l_1$ and $l_2$ be the magnitude of distances between the microphone 1 (M1) and source, and the microphone 2 (M2) and source respectively. Therefore, the phase difference between the sound received by M1 and M2 is given by,

$$\phi_{ij} = \frac{2\pi f}{c}(l_i - l_j) \qquad (10)$$

Hence, for $N$ microphones, using conventional beamforming, we obtain a ($N \times N$) matrix of phase differences.

*2.2.1 Far field approach – Infinite source*

In this approach, the sound source is assumed to be spread on the surface of a large hemispherical dome, with its center coinciding with the center of the microphone array disk.

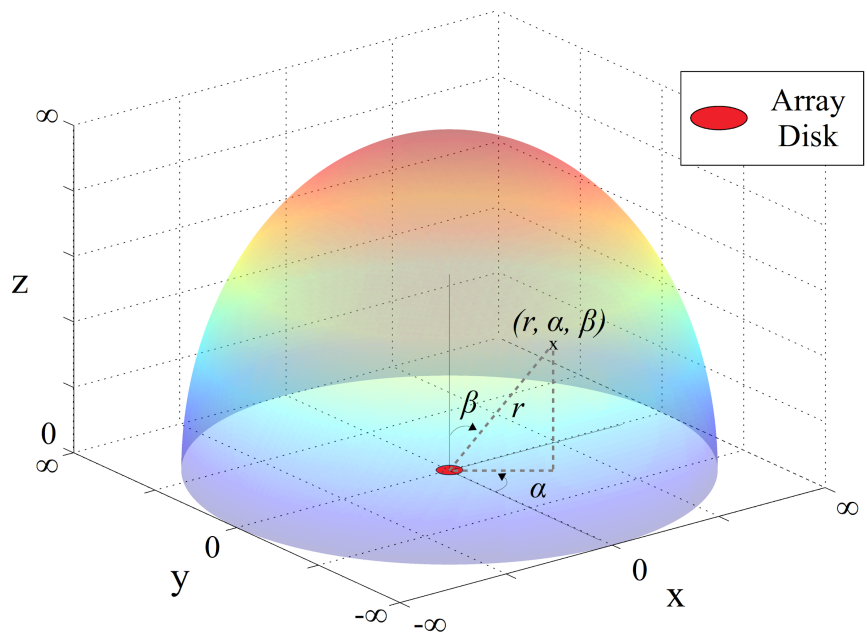

**Fig. 4.** Infinite dome sound source, and planar array.

As the source dome is infinite, theoretically there must exist infinite sound sources distributed on its surface. However, in our study, as the response is studied using a discretized approach, there exist finite source points equally spread on the surface of the dome. Using spherical coordinates, each source point can be represented using (r, α, β) coordinates. As $r \to \infty$, the two coordinates, polar angle (α) and azimuth angle (β) are sufficient to represent the sound source points.

The noise captured and corresponding rejection factor, using conventional beamforming, can be expressed as,

$$v_{noise}(\alpha,\beta) = \frac{1}{N^2}\sum_{i=1}^{N}\sum_{j=1}^{N}\left[v\left(x + \phi_{ij}(\alpha,\beta) - \phi_{o_{ij}}(\alpha,\beta)\right)\right] \quad (11)$$

$$RF = 1 - \left[\frac{1}{\pi^2}\int_{\alpha=-\frac{\pi}{2}}^{\alpha=\frac{\pi}{2}}\int_{\beta=-\frac{\pi}{2}}^{\beta=\frac{\pi}{2}}\frac{|v_{noise}(\alpha,\beta)|}{|v_{signal}|}(d\alpha)(d\beta)\right] \quad (12)$$

### 2.2.2 Near field approach – Finite source

In this approach, the sound source is assumed to be a finite disk of radius RS, but at a height (HS) above the microphone array disk.

Note: A finite disk source is taken for the purpose of analysis. This approach can be applied to a source of any arbitrary shape and size.

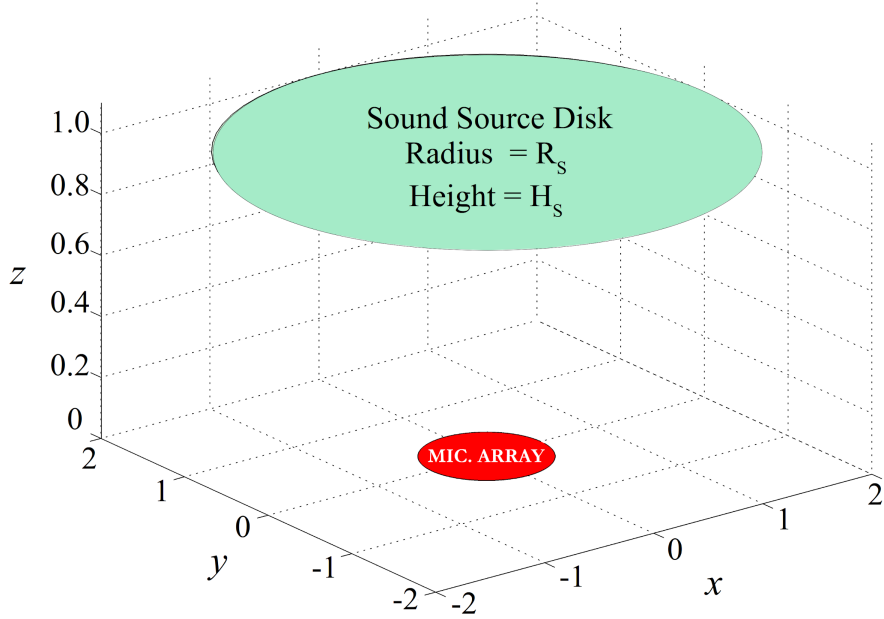

**Fig. 5.** Finite disk sound source, and planar array.

In this case too, the sound source is discretized into finite points. Cartesian coordinates are used to represent the sound source points. Hence, three parameters are required $-(x_s, y_s, H_s)$ coordinates. The noise captured and corresponding rejection factor, using conventional beamforming, can be expressed as,

$$v_{noise}(x_s, y_s, H_s) = \frac{1}{N^2}\sum_{i=1}^{N}\sum_{j=1}^{N}\left[v\left(x + \phi_{ij}(x_s, y_s, H_s) - \phi_{o_{ij}}(x_s, y_s, H_s)\right)\right] \quad (13)$$

$$RF = 1 - \left[ \frac{1}{\pi R_S^2} \int_{x_s=-R_S}^{x_s=R_S} \int_{y_s=-\sqrt{R_S^2-x_s}}^{y_s=\sqrt{R_S^2-x_s}} \frac{|v_{noise}(x_s,y_s,H_s)|}{|v_{signal}|} (dy_s)(dx_s) \right] \quad (14)$$

The higher the RF and higher the SNR, the better is the array performance. In 2D planar array conventional beamforming, it is not required to assume the sound waves to be planar waves (as in 1-D beamforming). This is due to the fact that the sound source is discretized, and distance between every microphone and each sound source is accounted for.

## 2.3 Grid Convergence

In order to assess the stability of the beamforming techniques, 'grid convergence' tests are carried out. 'Grid convergence' in this context means, the variation of **RF** as **dθ** (angular displacement of the sound source discretization) is progressively reduced. The beamforming scheme is stable if the value of RF converges to a constant value as we decrease $d\theta$.

Fig. 6 shows the convergence of the rejection percentage for a 1-D linear array, with input configuration as $\theta_o = 0°, f = 150 Hz, d = 0.2m, N = 4$ $and$ $c = 343 m/s$. A convergence criterion of $\varepsilon = (RF_{i+1} - RF_i)/RF_i = 10^{-3}$ is used, from which we obtain that convergence is achieved for $d\theta \approx 10^{-2} = 0.01°$.

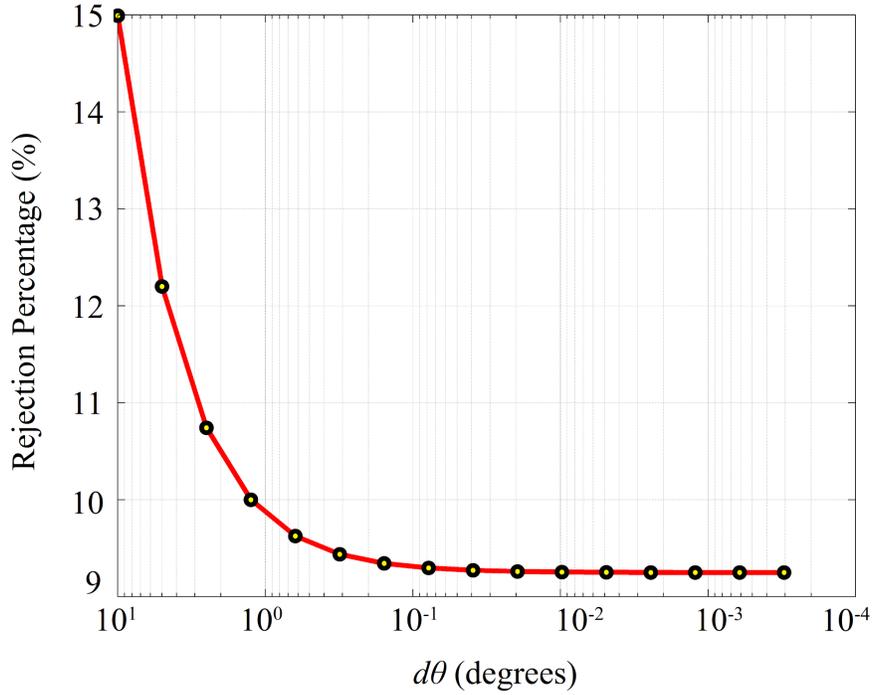

**Fig. 6.** Convergence of *RP* (%) for a linear array.

A similar convergence behavior is obtained for 2-D arrays as well, as shown in Fig. 7 and 8, using input array configuration of an Equi-area array as shown in Table 1, and sound source configuration of $f = 800Hz$ $and$ $c = 343m/s$. For 2-D array grid convergence tests, the step in $\alpha$ (polar angle) is taken to be equal to the step in $\beta$ (azimuth angle) i.e., $d\alpha = d\beta = d\theta$ (far-field), and the step in $x_s$ is taken to be equal to the step in $y_s$ i.e., $dx_s = dy_s = ds$ (near-field).

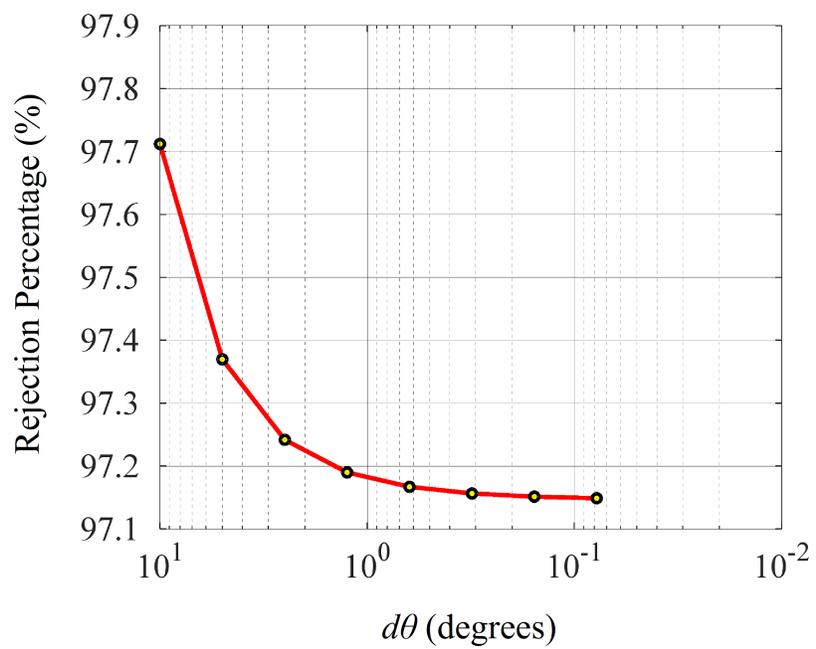

**Fig. 7.** Convergence of *RP* (%) for a 2-D array in far-field.

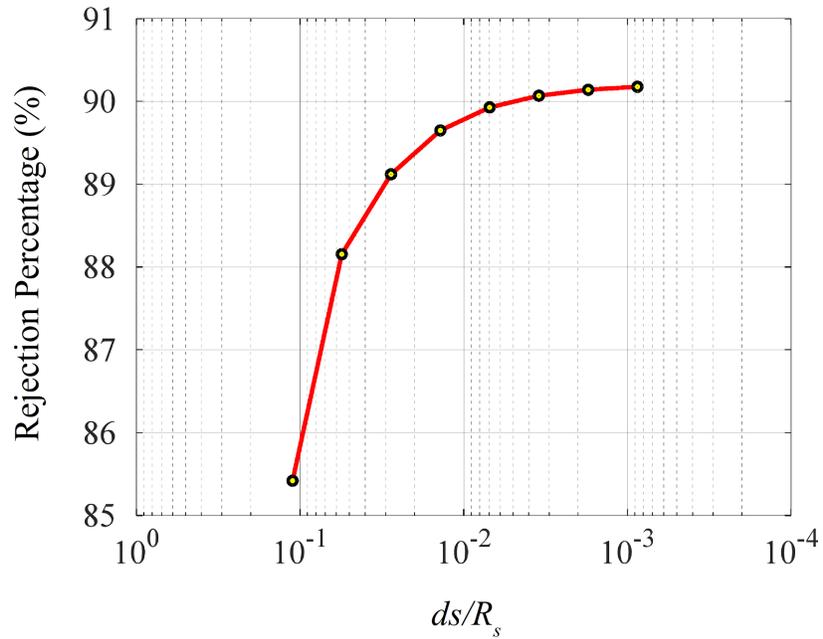

**Fig. 8.** Convergence of *RP* (%) for a 2-D array in near-field.

## 3. Analysis and Results

In this section, the parameters on which the Rejection factor (*RF*) depends on are analyzed. The incoming sound source is assumed to be a pure sound/tone, i.e., it consists of a single frequency (*f*), and hence the waveform can be modelled as $v(t) = sin(2\pi f.t)$.

## 3.1     1-D Linear Array

For the 1-D Linear array it is found that $RF$ depends on four parameters, N – number of microphones, d – microphone spacing, f – sound source frequency, c – speed of sound. It is found that it is possible to define a non-dimensional parameter $G$ such that the $RF$ remains constant for a constant $G$. For the 1-D linear array, $G$ takes the form:

$$G = \frac{Nfd}{c} \qquad (15)$$

This can be seen in Fig. 9, where the polar response of the array is shown. (It is desired to receive sound from 0°). Between each of these plots, only two parameters are varied. The similarity between the plots is clearly visible.

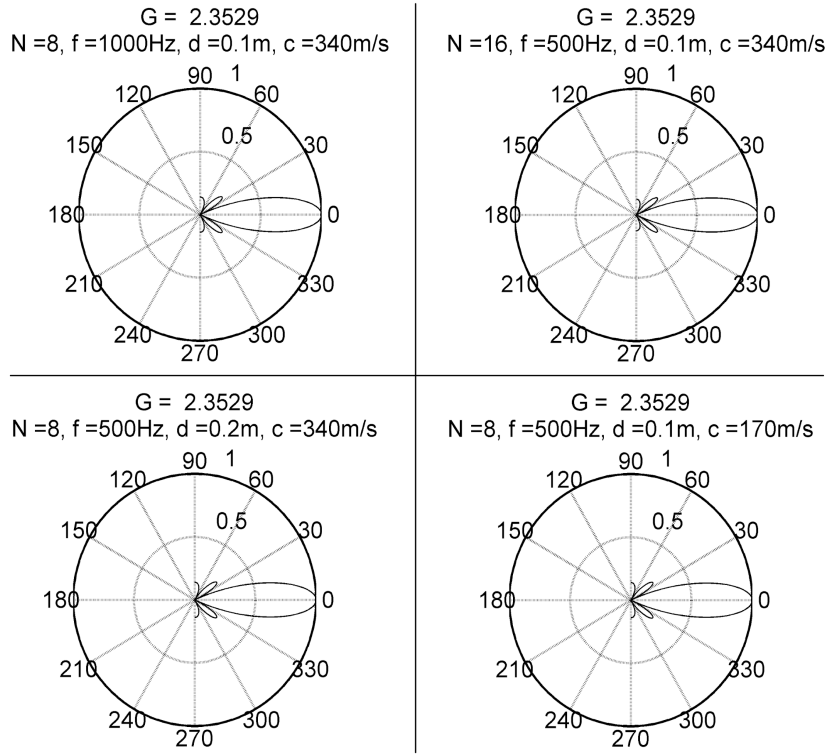

**Fig. 9.** Similar polar response for constant *G*.

It is interesting to note that the parameter *G* is a function of both the array design parameters and the properties of the sound source. This leads to the observation that for any given wavelength of incoming sound source ($\lambda = f/c$), it is possible to design a microphone array that will result in the required amount of noise rejection, by varying either the number of microphones or the spacing between microphones. This further leads us to the question of what is the highest amount of noise rejection that can be obtained with the *delay and sum beamforming* technique for 1-D linear arrays?

### 3.1.1 G-optimization

The next objective is to find the optimum $G$ for which $RF$ is maximized. In Fig. 10, the rejection percentage is plotted for increasing $G$.

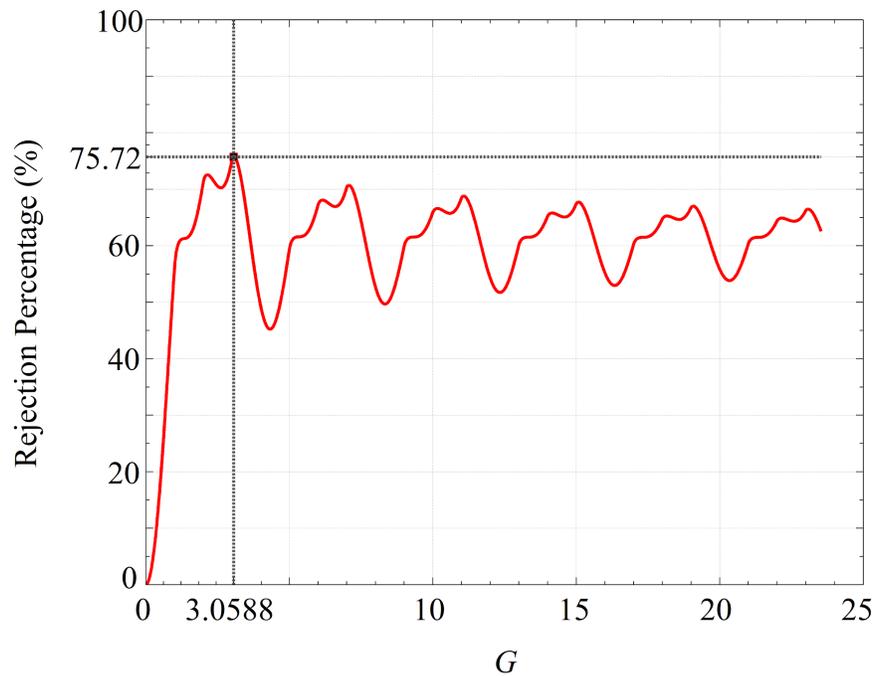

**Fig. 10.** Variation of RP (%) with $G$.

From the numerical simulations, it is found that the optimum *G* is 3.0588 and the maximum rejection percentage that can be obtained from a linear array is 75.72%. Looking at the *G* points where other peaks are obtained, it is found that such points of *G* produce side lobes, i.e., they receive sound from certain directions other than the desired direction equally efficiently as they receive it from the desired direction. This can be observed from the polar plots at these *G* points.

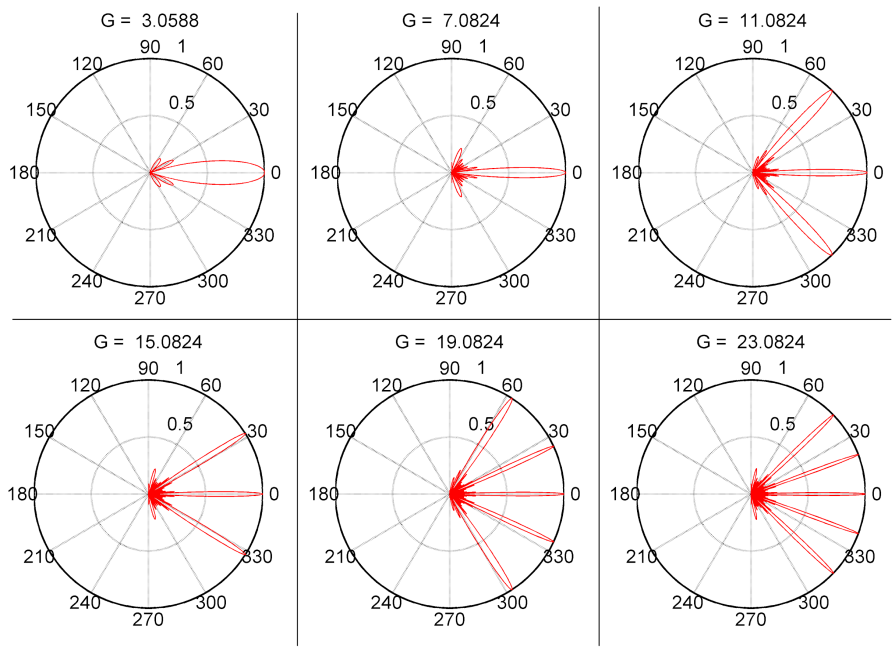

**Fig. 11.** Formation of side lobes after certain *G* .

In Fig. 11, the desired reception is from 0º. An important feature to notice is that these *G* points are equally spaced with a spacing of four units (~ 7, 11, 15, 19, 23) . Hence from the above study, we can conclude the feasible range of operation for a linear phased microphone array is for **$G \in [0,8]$**. Going above this *G* would introduce false results due to the generation of side lobes. Another disadvantage of linear arrays is the cap on the rejection percentage. Even with maximum design efficiency, the highest rejection that a linear array can achieve is 75.72%. It is due to this fact that the arrays are extended to a plane (instead of a straight line).

### *3.2    2-D Planar Array*

2-D array design is a non-trivial process as infinite possibilities and designs exist [2]. In this study, we consider a few popular designs and a new developed design, all comparing a total of 16 sensors. The different designs studied are 1) Concentric array, 2) Four arm spiral [16], 3) Archimedean spiral [17], 4) Underbrink array [18], 5) Equi-area array.

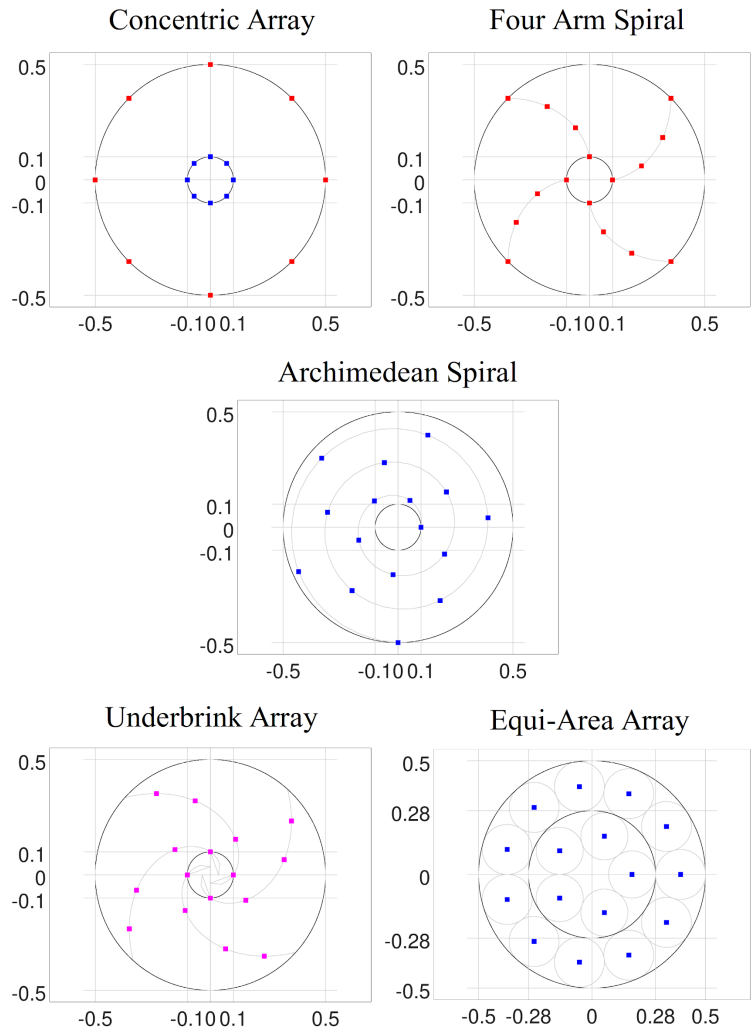

**Fig. 12.** Types of planar arrays analyzed.

1) **Concentric Array**

The microphones are spread equally into two concentric circles. The radii of the circles is $R_1$ and $R_2$, with $R_2 > R_1$.

### 2) Four-arm Spiral

The microphones are divided equally into 4-arms. Each arm starts at radial distance $R_1$ (head) and ends at radial distance $R_2$ (tail). The heads are placed at angular positions ($\delta$) of $0^o, 90^o, 180^o, 270^o$ respectively. The tails are placed at an angular displacement of 45 degrees from the head i.e., at $45^o, 135^o, 225^o, 315^o$ respectively. The radial and angular position of any microphone in between the head and tail is given by,

$$\delta_i = \delta_{head} + \frac{45 \times (i-1)}{\left(\frac{N}{4}-1\right)} \; degrees \tag{16}$$

$$r_i = R_1 + \frac{(R_2 - R_1) \times (i-1)}{\left(\frac{N}{4}-1\right)} \; m \tag{17}$$

where, $i = 1, 2 \ldots \left(\frac{N}{4}\right)$, 1-> Head, N/4-> Tail

### 3) Archimedean Spiral

The Archimedean spiral (also known as the arithmetic spiral) array is an array with microphones placed according to the Archimedean spiral equation, Eq. 18:

$$r(\theta) = a + b\theta \tag{18}$$

The radial and angular position of any microphone is given by,

$$\delta_i = \frac{\varphi \times (i-1)}{(N-1)} \; degrees \tag{19}$$

$$r_i = R_1 + \frac{(R_2 - R_1) \times (i-1)}{(N-1)} \, m \tag{20}$$

The spiral starts at radial distance $R_1$ (head) and ends at radial distance $R_2$ (tail). The total turn angle of the spiral is $\varphi$ degrees.

4) **Underbrink array**

The Underbrink design is a modified multi-spiral design, where the microphones are placed in the center of equal area segments. The procedure for calculating the microphone locations is to select the maximum and minimum radii, $R_2$ and $R_1$, the number of spiral arms, $N_a$, the number of microphones per spiral $N_m$, and the spiral angle, $\vartheta$. The area of the array is then separated into $N_m - 1$ equal area annuli, which are further subdivided into equal area segments, with microphones placed at the center of these segments. Finally, an inner circle of microphones is added at $R_1$. The radial locations of the microphones are:

$$r(m, 1) = R_1, \; m = 1, 2 \ldots N_a \tag{21}$$

$$r(m, n) = \sqrt{\frac{2n-3}{2N_m - 3}} R_2, \; m = 1, \ldots N_a, n = 2, \ldots N_m \tag{22}$$

With the radii of the microphones known, the angles are calculated by placing each microphone along a log spiral and rotating the spiral around the origin so that there are $N_a$ spiral arms. Thus, the angular positions of the microphones are:

$$\delta(m,n) = \frac{180 \times \ln\left(\frac{r(m,n)}{R_1}\right)}{\pi \times \cot(\vartheta)} + \frac{360 \times (m-1)}{N_a} \; deg \qquad (23)$$

$$m = 1, \ldots N_a, \qquad n = 1, \ldots N_m$$

**5) Equi-area array**

In this array design, the microphones are placed into two concentric circles, just like in Array 1, such that they are circumscribed by circles of equal area. Essentially, they are divided into two rings, outer ring (OR) and inner ring (IR). Given a disk of radius R, number of microphones in outer ring $N_{OR}$, and total number of microphones N, we can define the following,

$$s_1 = \frac{R \sin\left(\frac{\pi}{N_{OR}}\right)}{\left(1+\sin\left(\frac{\pi}{N_{OR}}\right)\right)} \text{ and } s_2 = \frac{(R-2s_1)\sin\left(\frac{\pi}{(N-N_{OR})}\right)}{\left(1+\sin\left(\frac{\pi}{(N-N_{OR})}\right)\right)} \qquad (24)$$

The radial and angular positions of each microphone is given by

$$r_i = R_{OR} = R - s_1, i = 1, \ldots N_{OR} \qquad (25)$$

$$r_j = R_{IR} = R - 2s_1 - s_2, j = 1, \ldots (N - N_{OR}) \qquad (26)$$

$$\delta_i = \frac{360}{N_{OR}}(i-1), i = 1, \ldots N_{OR} \qquad (27)$$

$$\delta_j = \frac{360}{N_{IR}}(j-1), i = 1, \ldots (N - N_{OR}) \qquad (28)$$

### 3.2.1 Array Response

The performance (rejection percentage, signal to noise ratio) of each array is evaluated and compared, for both near and far field noise rejection. A set of array parameters are chosen to perform the study, with the number of microphones and array outer dimensions kept uniform across all the designs (16 and 0.5m respectively).

**Table 1 – Parameters for five different arrays analyzed**

| | |
|---|---|
| Concentric | $N = 16, R_1 = 0.1\ R_2 = 0.5$ |
| Four-arm spiral | $N = 16, R_1 = 0.1\ R_2 = 0.5$ |
| Archimedean spiral | $N = 16, R_1 = 0.1\ R_2 = 0.5,$ $\varphi = 90°$ |
| Underbrink | $N = 16, R_1 = 0.1\ R_2 = 0.5,$ $N_a = 4, N_m = 4, \vartheta = 5\pi/16$ |
| Equi-area | $N = 16, N_{OR} = 11, R = 0.5$ |

Array responses have been analysed for three different frequencies, maintaining the speed of sound to be uniform (c=343 m/s) across the analysis. For far-field analysis, a radius of hemispherical sound dome = 100 m is used, with the desired reception direction being $(\alpha_o, \beta_o) = (0°, 0°)$. For near-field analysis, a radius of disk, $R_S = 2$m and a normal distance between array and source, $H_S = 0.1$m is used, with the desired reception location being $(x_o, y_o, z_o) = (0, 0, H_S)m$.

Table 2. Results summary: 2-D Planar Array

| Array Type | $f = 400\ Hz$ | | $f = 800\ Hz$ | | $f = 1200\ Hz$ | |
|---|---|---|---|---|---|---|
| | RP (%) | SNR (dB) | RP (%) | SNR (dB) | RP (%) | SNR (dB) |
| **Far Field Array Response** | | | | | | |
| Concentric | 85.48 | 16.77 | 81.25 | 14.54 | 93.07 | 23.19 |
| Four-arm spiral | 86.00 | 17.08 | 93.69 | 24.00 | 95.27 | 26.50 |
| Archimedean spiral | 85.37 | 16.69 | 96.37 | 28.80 | 96.33 | 28.70 |
| Underbrink | 78.94 | 13.53 | 94.16 | 24.67 | 97.02 | 30.51 |
| Equi-area | **89.32** | **19.43** | **97.08** | **30.69** | **98.14** | **34.60** |
| **Near Field Array Response** | | | | | | |
| Concentric | 57.91 | 07.52 | 80.67 | 14.28 | 91.43 | 21.34 |
| Four-arm spiral | 74.59 | 11.90 | 85.34 | 16.68 | 89.13 | 19.28 |
| Archimedean spiral | 79.03 | 13.57 | 87.75 | 18.24 | 88.14 | 18.52 |
| Underbrink | 73.74 | 11.61 | 87.33 | 17.94 | 87.88 | 18.33 |
| Equi-area | **83.23** | **15.51** | **93.72** | **24.04** | **92.20** | **22.16** |

The results show that the Equi-area array performs better than the other array types, for both near and far field, for a wide range of frequencies. However, there is a significant reduction in the performance of each array for near-field sound when compared to far-field sound. An advantage of the equi-area array is that it does not produce any side-lobes for the given configuration, and equally rejects noise from all directions. Another important observation is that the rejection percentage for 2-D planar arrays is significantly higher compared to the maximum 75.72 % for linear arrays, using the conventional beamforming technique.

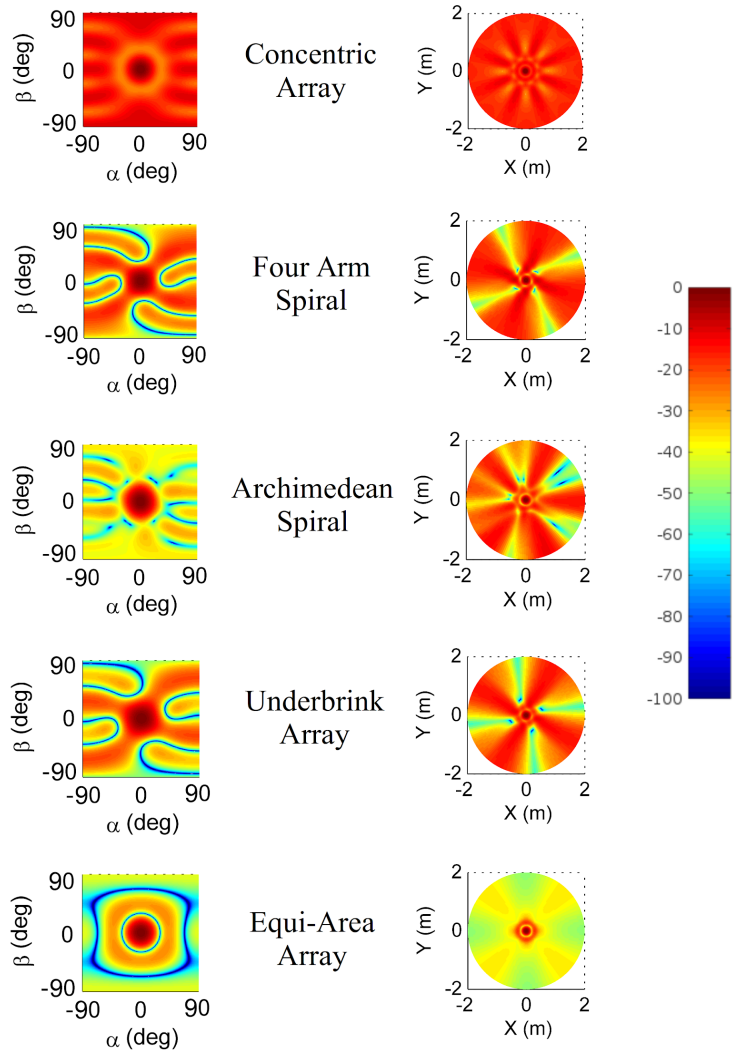

**Fig. 13.** Rejection Performance $v_{noise}/v_{signal}$ $(dB)$.

### 3.2.2 *Detailed study of the Equi-area array*

A parametric study is carried out, and similar to the linear array case, a non-dimensional parameter $G$ is defined for the equi-area array in *near field* response. The parameters that $G$ depends on are, $N, f, c, R, R_s$ and $H_s$. The expression for $G$ is obtained to be,

$$G = \frac{NfR}{c}\left(\frac{a+b.\log R_s}{\sqrt{1+d.H_s^2}}\right) \qquad (29)$$

From data fitting, array constants a, b and d are computed to be,

$$a = 1, b \approx 2, d \approx 34$$

Similar to the linear array case, the optimum $G$ for which $RF$ is maximized is evaluated. For this the rejection percentage is plotted for increasing $G$.

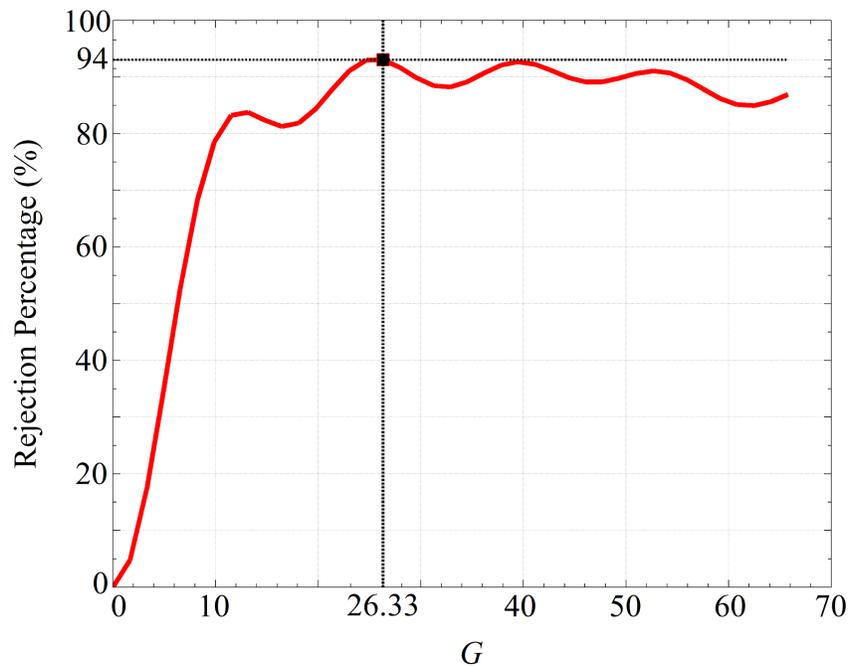

**Fig. 14.** Variation of RP (%) with $G$ for Equi-area array.

It is found that the optimum $G$ is 26.33 and the maximum rejection percentage that can be obtained from the equi-area array for near-field sound response is 94%.

## 4. Conclusions

- Different beamforming techniques (*delay and sum, conventional*) and their applications are discussed in detail.
- Performance parameters of a phased array microphone are defined, and performance of different array designs are quantified in terms of this performance parameter – Rejection Factor
- An upper limit to the Rejection factor is obtained for the linear array design (~ 75%) using the delay and sum beamforming technique.

- 2-D planar arrays are found to give better results compared to 1-D linear arrays with maximum rejection factors up to 97%, using the conventional beamforming technique.
- A new array design is proposed (Equi-area array), which is relatively simple to build. The performance of this array design is found to be exceeding that of other previously proposed designs, for both far-field and near-field sound sources.

5. **Sources of Errors**

- In the entire course of study, microphones are assumed to be point receivers. The size of microphones is not considered. Hence, the interference/blockage caused by one microphone on the others is not considered. This assumption is valid for large array systems, however it fails for small size arrays.
- All sound sources are considered to be coherent and point sources, i.e., they emit a single frequency from an infinitesimal point in space.
- The properties of the medium are assumed to be constant throughout (therefore c=constant).
- Effect of attenuation of sound with distance is not considered. The amplitude of sound waves is assumed to remain constant, with only its phase varying.